\documentclass[12pt]{article}
\usepackage{epsfig}
\begin{document}

\begin{titlepage}
\noindent
\begin{flushright}
March, 2005\\
MSUHEP-050310\\
hep-ph/0503098
\end{flushright}
\begin{center}
  \begin{large}
    \begin{bf}
        A Gauge-Mediation Model with a Light Gravitino of Mass
        O(10) eV and the Messenger Dark Matter
    \end{bf}
  \end{large}
\end{center}
\vspace{0.2cm}
\begin{center}
 Masahiro Ibe$^{(a)}$,~ Kazuhiro Tobe$^{(b)}$ and Tsutomu Yanagida$^{(a)(c)}$
\\
  \vspace{0.2cm}
  \begin{it}
    $^{(a)}$Department of Physics, University of Tokyo\\
    Tokyo 113-0033, Japan\\
    $^{(b)}$Department of Physics and Astronomy, 
    Michigan State University\\
    East Lansing, MI 48824, US\\
    $^{(c)}$Research Center for the Early Universe, University of Tokyo\\
    Tokyo 113-0033, Japan
  \end{it}
\end{center}

\begin{abstract}
In the light of recent experimental data on gaugino searches, we revisit the direct-transmission 
model of dynamical supersymmery breaking with the gravitino mass $m_{\tilde{G}}\leq 16$ eV, which
does not have any cosmological or astrophysical problems. We find that in the consistent regions of 
parameter
space, the model predicts not only upper bounds on superparticle masses ($1.1$ TeV, $320$ GeV, $160$ 
GeV, $5$ TeV, $1.5$ TeV and $700$ GeV for gluino, Wino, Bino, squarks, left-handed sleptons and 
right-handed sleptons, respectively), but also a mass of the lightest messenger particle 
in the range of $10-50$ TeV.
The lightest messenger particle can naturally be a messenger sneutrino.
Therefore, this may suggest that the messenger sneutrino could be the dark matter, as proposed 
recently by Hooper and March-Russel to account for the gamma-ray spectrum from the galactic center
observed by HESS experiment.
\end{abstract}
\end{titlepage}

A light gravitino of mass $\leq 16$ eV is very interesting, since it
does not cause any cosmological or astrophysical problems~\cite{Viel:2005qj}.  
In particular, we have no so-called gravitino problem~\cite{Weinberg:1982zq, Kawasaki:2004yh},
since the next lightest supersymmery (SUSY) particle decays sufficiently before
the Big Bang nucleosynthesis. Thus, the reheating temperature
after inflation can be higher than $10^{10}$ GeV, which is the lowest temperature
for the thermal leptogenesis to work~\cite{Fukugita:1986hr,Buchmuller:2005eh}. 
However, it is very difficult to construct
a consistent gauge-mediation model with such a light gravitino and only a quite few 
examples
are known~\cite{Izawa:1997hu,Izawa:1997gs,Izawa:1999vc,Izawa:2005yf}. 
In this letter, we discuss the model in Ref.~\cite{Izawa:1997gs} in the light of 
recent experimental data on the search for gauginos in a class of 
gauge-mediation models, and show that the minimal model in Ref.~\cite{Izawa:1997gs} 
is already excluded, but the next-to-minimal model
still survives. We show that the model predicts not only upper bounds on superparticle 
masses ($1.1$ TeV, $320$ GeV, $160$ GeV, $5$ TeV, $1.5$ TeV and $700$ GeV for gluino, Wino, Bino, 
squarks, left-handed sleptons and right-handed sleptons, respectively),
but also a mass of the lightest messenger particle in the range of $10-50$ TeV. This may 
suggest that the lightest messenger particle is the dark matter proposed recently 
by Hooper and March-Russell~\cite{Hooper:2004fh} to account for the gamma-ray spectrum 
from the galactic center~\cite{Aharonian:2004wa}.

Let us briefly describe a model discussed in Ref.~\cite{Izawa:1997gs}. We assume
a SUSY $SU(2)$ gauge theory with four doublet chiral superfields $Q^i_\alpha (\alpha = 1,2;
~ i=1-4)$ and six singlet chiral superfields $Z_{ij}=-Z_{ji}$~\cite{Izawa:1996pk,Intriligator:1996pu}. 
We impose, for simplicity, a flavor $SP(2)$ symmetry in the superpotential
\begin{equation}
W=\lambda_{ij}Q_{i}Q_jZ_{ij},
\end{equation}
and assume the $SP(2)$-invariant vacuum, $\langle Q_1Q_2 \rangle=\langle Q_3Q_4\rangle
=\Lambda^2$. Here, $\Lambda$
is the dynamical scale of the strong $SU(2)$ gauge interactions. Then, the effective
superpotential is given by
\begin{equation}
W_{eff}\simeq \lambda\Lambda^2Z.
\end{equation}
Here, the chiral superfield $Z$ is a $SP(2)$ singlet combination of $Z_{ij}$. We see 
that the $Z$ acquires a non-vanishing $F$ term ($F_Z\simeq \lambda\Lambda^2$) and the
SUSY is spontaneously broken~\cite{Izawa:1996pk,Intriligator:1996pu}.
 
The low energy effective superpotential in the messenger sector is given 
as follows~\cite{Izawa:1997gs}:
\begin{eqnarray}
W_{eff}&=& \lambda \Lambda^2 Z+
\sum_{a=1}^n \left\{Z(k_{d_a}d_a\bar{d}_a+k_{l_a}l_a\bar{l}_a)
+m_{d_a} d_a \bar{d'}_a + m_{\bar{d}_a} d'_a \bar{d}_a 
+m_{l_a} l_a \bar{l'}_a+m_{\bar{l}_a} l'_a \bar{l}_a \right\}.
\end{eqnarray}
Here we introduce $n$ sets of vector-like
messenger quark multiplets $d_a,\bar{d}_a,d'_a$ and $\bar{d'}_a$ and
lepton multiplets $l_a,\bar{l}_a,l'_a$ and $\bar{l'}_a$ ($a=1-n$).
We assume that the multiplets $(d,l)$ and $(d',l')$ 
transform as ${\bf 5}$, and $(\bar{d},\bar{l})$ and $(\bar{d}',\bar{l}')$
as ${\bf \bar{5}}$ under $SU(5)$, so that
the gauge coupling unification remains in the models.
Note that the perturbativity up to the GUT scale $(M_G=2\times 10^{16}~{\rm GeV})$ allows only cases with 
$n=1$ and $2$.\footnote{
Assuming $\alpha_3(m_Z)=0.12~(0.11)$ and $\Lambda=2.6\times 10^5$ GeV, 
we get $\alpha_3(M_G)\simeq 6~(1.1)$ at one-loop level in the model
with $n=3$. Therefore, the model with $n=3$ might be marginally allowed in some particular
parameter regions, but we do not discuss it further in this paper.}
We refer a case with $n=1$ as the minimal model and
a case with $n=2$ as the next-to-minimal model.
We also assume that the correction to Kahler potential for $Z$ field induces
the vacuum expectation value (vev) $\langle Z \rangle\simeq \Lambda$~\footnote{
The mass parameters $m_{d_a},m_{\bar{d}_a},m_{l_a}$ and $m_{\bar{l}_a}$
can be generated dynamically as discussed in Ref.~\cite{Izawa:1997gs}.
If $\langle Z \rangle=0$, we may introduce mass terms 
$\sum_a (M_{d_a} d_a \bar{d}_a+M_{l_a} l_a \bar{l}_a)$~\cite{Izawa:1997gs}.}.
Under the following condition,
\begin{eqnarray}
\left|m_{\psi_a}m_{\bar{\psi}_a}\right|^2 >
\left|k_{\psi_a} \langle F_Z\rangle \right|^2~~~(\psi=d~{\rm and}~l),
\end{eqnarray}
we find the SUSY-breaking vacuum is true one with vanishing vevs
of the messenger squarks and sleptons:
\begin{eqnarray}
\langle F_Z \rangle &\simeq& \lambda \Lambda^2,~
 \langle \psi_a \rangle=\langle \bar{\psi}_a \rangle
=\langle \psi'_a \rangle =\langle \bar{\psi}'_a \rangle=0~~~ (\psi=d~{\rm and}~l).
\end{eqnarray}

The mass terms of the messenger particles are represented as
\begin{eqnarray}
{\cal{L}} &=& -\sum_{a=1}^n \sum_{\psi=d,l} \left[
(\bar{\psi}_a,\bar{\psi'}_a )
M^{(\psi_a)} 
\left(
\begin{array}{c}
\psi_a\\
\psi'_a
\end{array}
\right)+{\rm h.c.}
+(\tilde{\psi}^*_a,\tilde{\psi}^{'*}_a,
\tilde{\bar{\psi}}_a,\tilde{\bar{\psi}}'_a)
\tilde{M}^{2(\psi_a)}
\left(
\begin{array}{c}
\tilde{\psi}_a\\
\tilde{\psi}'_a\\
\tilde{\bar{\psi}^*_a}\\
\tilde{\bar{\psi}^{'*}_a}
\end{array}
\right)\right],
\end{eqnarray}
where $\psi$ and $\tilde{\psi}$ denote fermionic and bosonic components
of the superfield $\psi$, respectively.
The mass matrices $M^{(\psi_a)}$ and  $\tilde{M}^{2(\psi_a)}$ are given by
\begin{eqnarray}
M^{(\psi_a)}&=&\left(
\begin{array}{cc}
m^{(\psi_a)} & m_{\bar{\psi_a}}\\
m_{\psi_a} & 0
\end{array}
\right),
\\
\tilde{M}^{2(\psi_a)} &=& \left(
\begin{array}{cccc}
\left|m^{(\psi_a)}\right|^2+\left|m_{\psi_a}\right|^2 & m^{(\psi_a)*}m_{\bar{\psi}_a}
& F^{(\psi_a)*} & 0\\
m^{(\psi_a)}m_{\bar{\psi}^*_a} & \left|m_{\bar{\psi}_a}\right|^2 &0 &0\\
F^{(\psi_a)} & 0 &\left|m^{(\psi_a)}\right|^2+\left|m_{\bar{\psi}_a}\right|^2 &
m^{(\psi_a)}m_{\psi_a}^* \\
0& 0 & m^{(\psi_a)*}m_{\psi_a} & \left|m_{\psi_a}\right|^2
\end{array}
\right).
\label{scalar_messenger_mass}
\end{eqnarray}
Here $m^{(\psi_a)}=k_{\psi_a} \langle Z \rangle$ and 
$F^{(\psi_a)}=k_{\psi_a} \langle F_Z \rangle$ $(\psi=d~{\rm and}~l)$.
Diagonalizing the mass matrices, we obtain masses of the messenger particles.

Once the messengers receive SUSY breaking masses, gaugino (sfermion) masses 
in the minimal SUSY standard model (MSSM) sector are 
generated by one-loop (two-loop) diagrams of the messenger 
particles via standard model gauge interactions.
The gaugino masses are given by
\begin{eqnarray}
m_{\tilde{g}_3} &=& \frac{\alpha_3}{2\pi} \sum_{a=1}^n {\cal F}^{(d_a)},\\
m_{\tilde{g}_2} &=& \frac{\alpha_2}{2\pi} \sum_{a=1}^n {\cal F}^{(l_a)},\\
m_{\tilde{g}_1} &=& \frac{\alpha_1}{2\pi} \sum_{a=1}^n 
\left(\frac{2}{5}{\cal F}^{(d_a)}+\frac{3}{5}{\cal F}^{(l_a)}
\right),
\end{eqnarray}
and the sfermion masses are 
\begin{eqnarray}
m^2_{\tilde{f}}&=&\frac{1}{2}\sum_{a=1}^n \left[
C_3^{f} \left(\frac{\alpha_3}{4\pi}\right)^2 {\cal G}^{(d_a)2}
+C_2^f \left(\frac{\alpha_2}{4\pi}\right)^2 {\cal G}^{(l_a)2}
+\frac{3}{5}Y^2 \left(\frac{\alpha_1}{4\pi}\right)^2
\left(\frac{2}{5} {\cal G}^{(d_a)2}+\frac{3}{5} {\cal G}^{(l_a)2}
\right) \right],
\end{eqnarray}
where we have adopted the $SU(5)$ GUT normalization of the $U(1)_Y$ gauge coupling
($\alpha_1=\frac{5}{3} \alpha_Y$), and 
$C_3^f=\frac{4}{3}$ and $C_2^f=\frac{3}{4}$ when $\tilde{f}$ is in the fundamental
representation of $SU(3)_C$ and $SU(2)_L$ respectively, and $C_{3,2}^f=0$ for 
the gauge singlets, and $Y$ denotes the $U(1)_Y$ hypercharge ($Y=Q-T_3$).
Here ${\cal F}^{(\psi)}$ and ${\cal G}^{(\psi)}$
are functions of the messenger masses and mixings, and their explicit
expressions can be found in Ref.~\cite{Izawa:1997gs}.

Since SUSY is broken, the gravitino gets a mass:
\begin{eqnarray}
m_{\tilde{G}}=\frac{F_Z}{\sqrt{3} M_*}
=16 \left(\frac{\sqrt{F_Z}}{2.6\times 10^5~{\rm GeV}} \right)^2~{\rm eV}.
\label{gravitino_mass}
\end{eqnarray}
Here $M_*$ is the reduced Planck mass ($M_*=2.4\times 10^{18}$ GeV).
As pointed out in Ref.~\cite{Viel:2005qj}, the matter power
spectrum inferred from large samples of Lyman-$\alpha$ forest data and 
the cosmic microwave background data of WMAP strongly constrain the gravitino mass.
As a result, its current upper limit is 16 eV.\footnote{
To evade the bound on the gravitino mass, one needs to consider 
a late-time entropy production, as suggested in Ref.~\cite{Fujii:2002fv}.}
From Eq.~(\ref{gravitino_mass}), the gravitino mass limit translates into a limit on the 
SUSY breaking scale:
\begin{eqnarray}
\sqrt{F_Z}<2.6\times 10^5~{\rm GeV} \longleftrightarrow m_{\tilde{G}} \leq 16~{\rm eV}.
\label{SUSY_limit}
\end{eqnarray}
Note that in the model discussed here, the gravitino mass and SUSY breaking vev $F_Z$ are 
related each other as shown in Eq.~(\ref{gravitino_mass}), because the SUSY breaking in the dynamical
SUSY breaking sector is directly transmitted to the messenger sector.
In most of gauge-mediation models in literatures~\cite{Giudice:1998bp}, 
however, the SUSY breaking scale
in the messenger sector is suppressed, compared to the original
SUSY breaking scale, due to the transmission mechanism. 
Therefore we stress that the gravitino mass limit is quite
severe in most of gauge-mediation models, and it is crucial to have the direct-transmission
of dynamical SUSY breaking in order to construct the gauge-mediation models
with $m_{\tilde{G}}\leq 16$ eV.

\begin{figure}
\centering
\includegraphics*[width=10cm,angle=0]{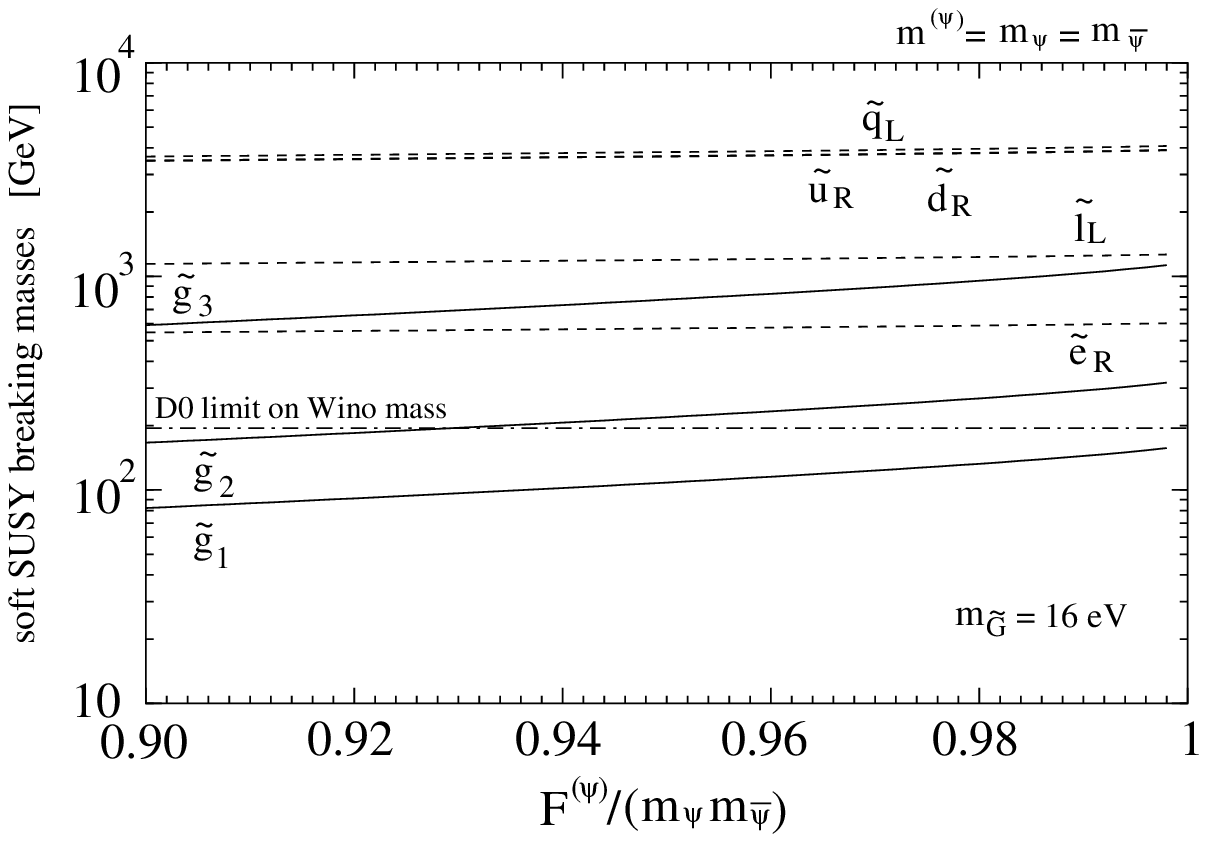}
\caption{Mass spectrum of the gauginos (solid lines) and sfermions (dashed lines) in the MSSM sector
 as a function of a parameter $F^{(\psi)}/(m_\psi m_{\bar{\psi}})$
in the next-to-minimal model.
Here we have assumed $m^{(\psi)}=m_{\bar{\psi}}=m_{\psi}$ for $\psi=d,~l$.
We also fixed $F^{(\psi)}=F_Z=(2.6\times 10^5~{\rm GeV})^2$ 
for $\psi=d,~l$, which corresponds to be the maximal
gravitino mass in Eq.~(\ref{SUSY_limit}), $m_{\tilde{G}}=16$ eV. The D0 limit on Wino mass (195 GeV)
is also shown.}
\label{SUSY_mass_limits}
\includegraphics*[width=10cm,angle=0]{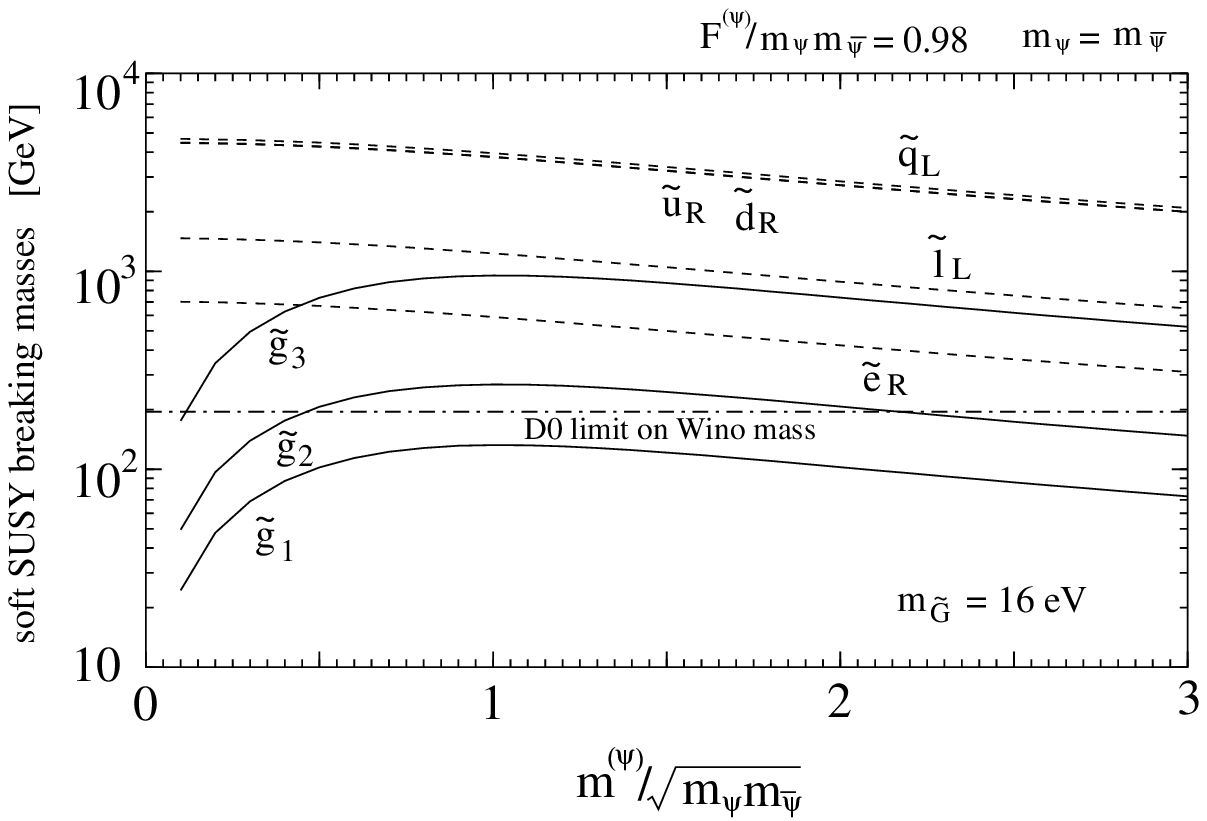}
\caption{Mass spectrum of the gauginos (solid lines) and sfermions (dashed lines) in the MSSM sector
 as a function of a parameter $m^{(\psi)}/\sqrt{m_\psi m_{\bar{\psi}}}$
in the next-to-minimal model.
Here we have assumed $F^{(\psi)}/(m_\psi m_{\bar{\psi}})=0.98$
and $m_{\bar{\psi}}=m_{\psi}$ for $\psi=d,~l$,
and $m_{\tilde{G}}=16$ eV. The D0 limit on Wino mass (195 GeV)
is also shown.}
\label{SUSY_mass_limits2}
\end{figure}

Now we are in position to discuss the prediction of the model based on
the limit in Eq.~(\ref{SUSY_limit}). Since the minimal model
(the model with $n=1$) has been excluded as we will see later, we mainly consider
the next-to-minimal model (the model with $n=2$).
In Fig.~\ref{SUSY_mass_limits}, mass spectrum of the gauginos and sfermions in the MSSM sector
are plotted as a function of $F^{(\psi)}/(m_\psi m_{\bar{\psi}})$ in the case of
the next-to-minimal model.\footnote{Within a particle content of MSSM, $\mu$-term tends to be
larger than Wino mass because squarks are much heavier than the Wino, as discussed 
in Ref.~\cite{Izawa:1997gs}. Therefore, Higgsinos are heavier than Winos.
However, the Higgs sector can be modified, as we will discuss later, and hence
it will be model-dependent. Thus we will not discuss the Higgs and Higgsino sector in detail here.
A more detail analysis will be given in Ref.~\cite{ITY}.}
Throughout our discussion, we assume $F^{(\psi_a)}\equiv F^{(\psi)}$, $m^{(\psi_a)}\equiv m^{(\psi)}$,
$m_{\psi_a}\equiv m_{\psi}$, and $m_{\bar{\psi}_a}\equiv m_{\bar{\psi}}$ for $a=1,2$, and hence
we suppress the index $a$, for simplicity. In Fig.~\ref{SUSY_mass_limits},
we have assumed
all F-parameters\footnote{
In our results, we assume the Yukawa couplings $k_{d_a}=k_{l_a}=1$ because we expect
these Yukawa couplings are of order one. One can easily estimate the change of our 
results when these Yukawa couplings are deviated from one.}
are equal, $F^{(l)}=F^{(d)}=F_Z=(2.6\times 10^5~{\rm GeV})^2$, which corresponds to $m_{\tilde{G}}=16$ eV,
and all mass parameters are equal, $m^{(\psi)}=m_{\psi}=m_{\bar{\psi}}$ for $\psi=d,~l$. 
In Fig.~\ref{SUSY_mass_limits2}, we also show the mass spectrum of the gauginos and sfermions
as a function of a parameter $m^{(\psi)}/\sqrt{m_\psi m_{\bar{\psi}}}$.
Here we assume $m_\psi = m_{\bar{\psi}}$ and $F^{(\psi)}/(m_\psi m_{\bar{\psi}})=0.98$
as an example. We see that the gaugino masses are maximal for $m^{(\psi)}/\sqrt{m_\psi m_{\bar{\psi}}}\simeq 1$ 
with fixed $F^{(\psi)}/(m_\psi m_{\bar{\psi}})$.
Because of the limit in Eq.~(\ref{SUSY_limit}), we find, in Fig.~\ref{SUSY_mass_limits}, 
the next-to-minimal model predicts upper limits on
superparticle masses ($1.1$ TeV, $320$ GeV, $160$ GeV, 
$5$ TeV, $1.5$ TeV and $700$ GeV for gluino, Wino, Bino, squarks, left-handed sleptons and 
right-handed sleptons, respectively).

Important experimental bounds on gaugino masses have been set by D0 and CDF 
experiments~\cite{Abazov:2004jx,Acosta:2004sb}.\footnote{LEP experiments also have some constraints
on gaugino masses for gauge-mediation models. However, in the models considered here,
sleptons are so heavy that their limits do not give a significant constraint.}
They have been searching for diphoton events
induced by the lightest neutralino decay into a gravitino plus a photon,
subsequent to Wino pair production at Tevatron. 
The D0 lower limit on Wino mass is about 195 GeV~\cite{Abazov:2004jx} and 
currently it is the strongest bound for the models
considered here. In Figs.~\ref{SUSY_mass_limits} and \ref{SUSY_mass_limits2},
the D0 limit is also shown. 

Note that in the minimal model ($n=1$), the predicted upper limits on gaugino
masses are half of those in the next-to-minimal model shown in 
Fig.~\ref{SUSY_mass_limits}, and hence the predicted Wino mass in the minimal model is 
smaller than about 160 GeV. Therefore, the D0 bound has
excluded the minimal model. Since models with $n \geq 3$
are not allowed by the perturbativity up to the GUT scale as we have mentioned before,
the next-to-minimal model is the only viable model.

The D0 bound constrains the parameter space in the next-to-minimal model.
For example, the parameter $F^{(l)}/(m_l m_{\bar{l}})$ has to be larger than about 0.93
for $m^{(l)}/\sqrt{m_l m_{\bar{l}}}$=1,
and $m^{(l)}/\sqrt{m_l m_{\bar{l}}}$ should be in the range of 
$0.5-2$ for $F^{(l)}/(m_l m_{\bar{l}})=0.98$, 
as shown in Figs.~\ref{SUSY_mass_limits} and~\ref{SUSY_mass_limits2}.
The experimental bound on gluino mass is about $200$ GeV for the next-to-minimal model, 
and hence a constraint on $F^{(d)}/(m_d m_{\bar{d}})$ is weaker than that of 
$F^{(l)}/(m_l m_{\bar{l}})$.

\begin{figure}
\centering
\includegraphics*[width=10cm,angle=0]{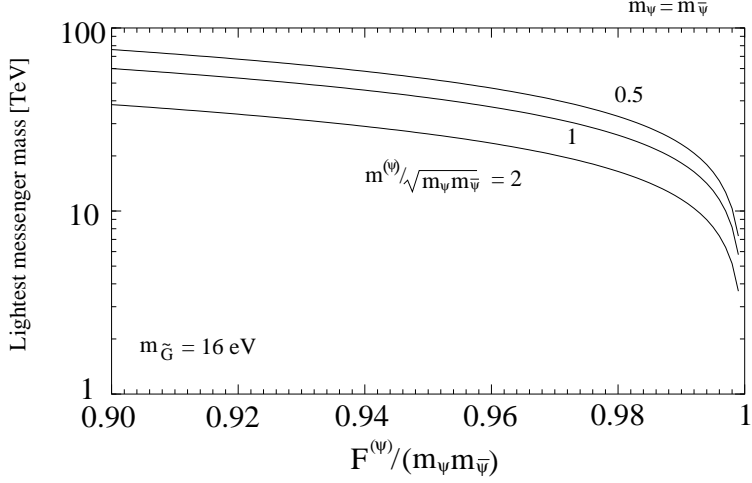}
\caption{The lightest messenger mass as a function of a parameter 
$F^{\psi}/(m_\psi m_{\bar{\psi}})$. We show lines for 
$m^{(\psi)}/\sqrt{m_{\psi} m_{\bar{\psi}}}=0.5,~1$ and $2$.
Here we have assumed that $m_{\psi}=m_{\bar{\psi}}$
and $m_{\tilde{G}}=16$ eV.}
\label{messenger_mass_limit}
\end{figure}

We find that there is an interesting consequence in the consistent region with
$F^{(l)}/(m_l m_{\bar{l}}) \sim 1$ and $m^{(l)}/\sqrt{m_l m_{\bar{l}}}\sim 1$.
In the region with $F^{(l)}/(m_l m_{\bar{l}}) \sim 1$ and $m^{(l)}/\sqrt{m_l m_{\bar{l}}}\sim 1$,
one of messenger sleptons gets lighter and it becomes the lightest messenger particle,
provided that $F^{(d)}/(m_d m_{\bar{d}})<F^{(l)}/(m_l m_{\bar{l}})$.
If $F^{(l)}/(m_l m_{\bar{l}})=1-\delta~(\delta\ll 1$), the lightest messenger slepton mass
is approximately given by
\begin{eqnarray}
m_{\tilde{\psi}_l}\simeq m\sqrt{\frac{\delta}{2}},
\end{eqnarray}
which can be calculated by the diagonalization of the mass matrix in 
Eq.~(\ref{scalar_messenger_mass}). 
Here we have assumed that all mass parameters are equal to $m$,
$m^{(l)}=m_{l}=m_{\bar{l}}\equiv m$.
If $\sqrt{F^{(l)}}\leq 2.6\times 10^5$ GeV (in other words, $m_{\tilde{G}}\leq 16$ eV)
and $F^{(l)}/(m_l m_{\bar{l}})\geq 0.93$,
the lightest messenger slepton mass should be smaller than about 50 TeV.
In Fig.~\ref{messenger_mass_limit}, we show a numerical result of the lightest messenger 
mass as a function of $F^{(\psi)}/(m_\psi m_{\bar{\psi}})$. 
Fig.~\ref{messenger_mass_limit} also shows the dependence on $m^{(\psi)}/
\sqrt{m_\psi m_{\bar{\psi}}}$.
Here we have assumed $m_{\tilde{G}}=16$ eV, and 
$m_\psi=m_{\bar{\psi}}$. 
From Fig.~\ref{messenger_mass_limit}, we see the light messenger slepton
mass in the range of about $10-50$ TeV is predicted in the consistent region of the next-to-minimal model.

The messenger scalar masses within the same $SU(2)$ doublet slepton multiplet are split 
at tree level by $SU(2)_L\times U(1)_Y$ D-terms. 
However the tree-level splitting is negligible~\cite{Dimopoulos:1996gy}
in the models studied here. 
The important splitting is generated at one-loop level by gauge boson loops
as follows~\cite{Han:1997wn}:
\begin{eqnarray}
m_{\tilde{\psi}_e}-m_{\tilde{\psi}_\nu} \simeq \frac{1}{2}\alpha m_Z,
\end{eqnarray}
where $m_{\tilde{\psi}_e}$ ($m_{\tilde{\psi}_\nu}$) is a mass of
a charged (neutral) component of the messenger doublet slepton. Therefore
the neutral component of the messenger slepton, which we call as the messenger sneutrino, 
can naturally be the lightest messenger particle, and it can be stable in the model.
Interestingly, the stable lightest messenger sneutrino
would be a dark matter candidate
if its mass is in the range of $10-30$ TeV as pointed out in 
Refs.~\cite{Han:1999jc,Hooper:2004fh}.
In order to make a viable model for the messenger dark matter, 
we introduce an extra singlet
$N$ and the following additional superpotential~\cite{Han:1999jc}:
\footnote{The superpotential in this model is consistent with R-symmetry, and hence it
is natural.}
\begin{eqnarray}
W=\sum_{\psi=d,l} \lambda_\psi N \bar{\psi}_1 \psi_2 +\lambda_H N H_u H_d +\frac{\lambda_N}{3}N^3.
\end{eqnarray}
The interaction of the singlet $N$ could increase the annihilation cross-section of
the messenger sneutrino $\langle \sigma v \rangle$ and it would provide a thermal relic density
of the messenger sneutrino $\Omega_{\tilde{\psi}_\nu} h^2$  consistent with the
measured dark matter density~\cite{Han:1999jc,Hooper:2004fh}:
\begin{eqnarray}
\nonumber 
\Omega_{\tilde{\psi}_\nu} h^2 &\simeq& 0.1 \times
\left(\frac{10^{-26}~{\rm cm}^3/s}{\langle \sigma v \rangle}\right),\\
\langle \sigma v \rangle &\simeq& \frac{y^4 \Lambda^4}{32 \pi m^6_{\tilde{\psi}_\nu}}
\sim 10^{-26} ~{\rm cm}^3/s~\times \left(\frac{y}{0.4}\right)^4
\left(\frac{\Lambda}{2.6\times 10^5~{\rm GeV}} \right)^4
\left(\frac{30~{\rm TeV}}{m_{\tilde{\psi}_\nu}}\right)^6,
\end{eqnarray}
where $y$ is a function of Yukawa couplings $k_{\psi_a}, \lambda_\psi,~\lambda_H$ and $\lambda_N$, 
which is of order one.

Furthermore, recently Hooper and March-Russell~\cite{Hooper:2004fh} 
has proposed that if the mass of 
messenger sneutrino dark matter is about $20-30$ TeV, it could also account for
the multi-TeV gamma-ray spectrum from the galactic center
observed by HESS~\cite{Aharonian:2004wa, Horns:2004bk}. As can be seen from 
Fig.~\ref{messenger_mass_limit}, the next-to-minimal model can predict such a messenger dark matter
if $F^{(l)}/(m_l m_{\bar{l}})$ is about $0.97-0.99$. In this range of parameter space,
the predicted Wino mass is about $250-290$ GeV. The gluino (Bino) is also predicted
to be lighter than $1$ TeV ($280$ GeV) since $F^{(d)}/(m_d m_{\bar{d}}) <F^{(l)}/(m_l m_{\bar{l}})$ 
in order for the messenger sneutrino to be lighter than the messenger squarks.\footnote{
We note that if we assume the unification of mass parameters between messenger quarks and leptons 
at the GUT scale, we get a relation: $F^{(l)}/(m_l m_{\bar{l}}) \simeq 2 F^{(d)}/(m_d m_{\bar{d}})$. 
However, this relation can be easily
changed by the GUT threshold corrections~\cite{Izawa:1997gs}. Also if the messenger quarks and leptons
belong to different GUT multiplets, there is no such a unification.}

In this paper, we have discussed the direct-transmission model of dynamical SUSY breaking proposed
in Ref.~\cite{Izawa:1997gs}. We have found that in order to be consistent with the current 
cosmological bound on the gravitino mass ($m_{\tilde{G}}<16$ eV), the direct-transmission of 
SUSY breaking is required. 
Combined with the current D0 limit on Wino mass,
the minimal model has been ruled out, and the next-to-minimal model is the only consistent
model. We have shown that predictions of upper limits on superparticle masses 
($1.1$ TeV, $320$ GeV, $160$ GeV, $5$ TeV, $1.5$ TeV and $700$ GeV for gluino, Wino, Bino, 
squarks, left-handed sleptons and 
right-handed sleptons, respectively).
We have also found that in the consistent region,
the next-to-minimal model predicts the light messenger slepton in the mass range 
of $10-50$ TeV, and hence
it would be an interesting dark matter candidate to account for the multi-TeV gamma-ray spectrum
observed by HESS experiment. Therefore, not only current and future sparticle searches but also more
multi-TeV gamma-ray data from HESS and Cangaroo III in the coming years would provide a further important 
insight for the direct-transmission model of the dynamical SUSY breaking.

\section*{Acknowledgements}
K.T. thanks Wayne Repko for a careful reading of the manuscript.
K.T. acknowledges support from the National Science Foundation.

\end{document}